\documentclass[twoside,a4paper]{article}
\usepackage{fleqn,epsf,graphicx}
\usepackage{authblk}

\begin{document}

\title{Radiation damage effects on detectors and eletronic devices in harsh radiation environment}

\author{S. Fiore}

\affil{ENEA UTTMAT-IRR, via Anguillarese 301, Roma, Italy\\
       INFN sezione di Roma, p.le Aldo Moro 2, Roma, Italy}

\maketitle                   

{PACS: 07.89.+b, 07.87,+v, 95.55.-n, 29.40.-n, 87.55.d-}

\begin{abstract}
Radiation damage effects represent one of the limits for technologies to be used in harsh radiation environments as space, radiotherapy treatment, high-energy phisics colliders. Different technologies have known tolerances to different radiation fields and should be taken into account to avoid unexpected failures which may lead to unrecoverable damages to scientific missions or patient health.
\end{abstract}

\section{Introduction}
Electronic devices are  often exposed to exceptional radiation levels, in many different fields of operation. Advances in technology during the last decades have both brought electronics to high radiation natural environment, and created situations in which high radiation levels are delivered to general population.

Space missions, artificial satellites orbiting around the Earth, high atmosphere experiments are examples of technological devices brought to natural high radiation environments. Nuclear power plants create artificial radiation environments, where technology is used and has to be able to operate to prevent catastrophic disasters. 
Radiation therapy for cancer disease can expose human body to extremely intense ionizing radiation levels, based on the scientific knowledge that the benefits of treating an existing disease would be more than the risk to create new damages. Scientific research in High-Energy physics, using colliding particle beams, necessarily exposes radiation detectors and related electronics to ionizing and non-ionizing radiation, and asks for long-term operating capabilities in order to continue to take data for years without detector maintenance.
These examples explain how electronic devices are more and more required to be resistant to radiation damage effects.

The various types of damage which can occur due to exposure to photons, high energy charged and neutral particles, slow neutrons, have been categorized as follows:
\begin{itemize}
\item effects due to ionizing radiation: these are due to the accumulation of the resulting free charges which can modify the properties of materials, and thus change the behavior of electronic components.
Under this category, electronic devices are known to suffer from Total Ionizing Dose (TID) effects and Single Event Effects (SEE). TID is caused by the progressive accumulation of charges over prolonged exposure to ionizing radiation, and is also known as surface damage. It represents an important effect for insulators, electronic components, where surface charge can affect the behavior of transistors due to change in thin oxide layer properties, optical elements, whose light absorption spectra can be modified. SEE instead are due to single ionization events wherein a large concentrated ionization gives a temporary or permanent damage to electronically live devices or systems. It is an important effect for digital circuits such as memories or microprocessors, where it can induce errors, undesired latch$‐$ups and may lead to system failures.
\item Effects due to non-ionizing radiation: particles can displace atoms from their lattice sites and produce bulk damage effects called Displacement Damage Dose (DDD) effects, or also Non-Ionizing Energy Loss (NIEL) effects. These are cumulative effects due to long-term exposure to interactions with non$‐$ionizing energy transfers, which originate displacement defects in semiconductor materials. These are important effects in all semiconductor bulk$‐$based devices.
\end{itemize}

\section{Radiation damage in High Energy Physics experiments}

Detectors which collect data from particle collisions are exposed to increasing levels of radiation, depending on the signatures they are built to detect. Detectors looking for particle tracks close to the interaction region of the colliding beams are the most exposed to ionizing and non-ionizing radiation from secondary particles \cite{atlas,cmstdr,babar}, especially in hadronic colliders, and up to now the tracking detectors with the highest spatial resolution use silicon technology which is sensitive to both TID and DDD effects.
Ionizing energy loss generates charges that can get trapped in the oxide layers and interfaces of silicon strip detectors. This can affect the interstrip capacitance, increasing the detector noise levels, and also the breakdown voltage of the strips. On the other hand, DDD can create bulk damage to silicon crystalline structure increasing the leakage current, increasing the number of trapped charges, and changing the effecive doping concentration. This last effect can cause the so-called "type inversion", which changes the effective depth and growing direction of the depleted region of the silicon volume.
Typical levels of tolerance for silicon detectors are of the order of 10$^{13}$ hadrons/cm$^2$, with some improved hardness detectors tolerating up to 10$^{15}$ hardons/cm$^2$. 
Detectors making use of gas mixtures are also sensitive to TID effects. The ionization of the gas molecules, combined with the high electric field,  induces the formation of polymers which deposit around the anode wires and create a dielectric shielding where charges build up, reducing the electric field and consequently the gain and collection efficiency of the detector. Polymers can also accumulate on the cathode, causing the extraction of charges from it and increasing noise \cite{capeans}. The tolerated radiation levels differ dramatically for different detector layouts and gas mixtures, the micro pattern detectors being less sensitive to radiation due to their planar geometry and lower electric fields when used in cascade configuration. Orders of magnitude of 10$^{11}$ minimum ionizing particles per mm$^2$ can start causing ageing effects in gas detectors. Additional unwanted contamination of gas mixture can increase radiation effects, introducing  molecular species that can be ionized and create unpredicted dielectric layers \cite{capeans}. 

\begin{figure}[h!]
\includegraphics  [width = 0.8  \textwidth] {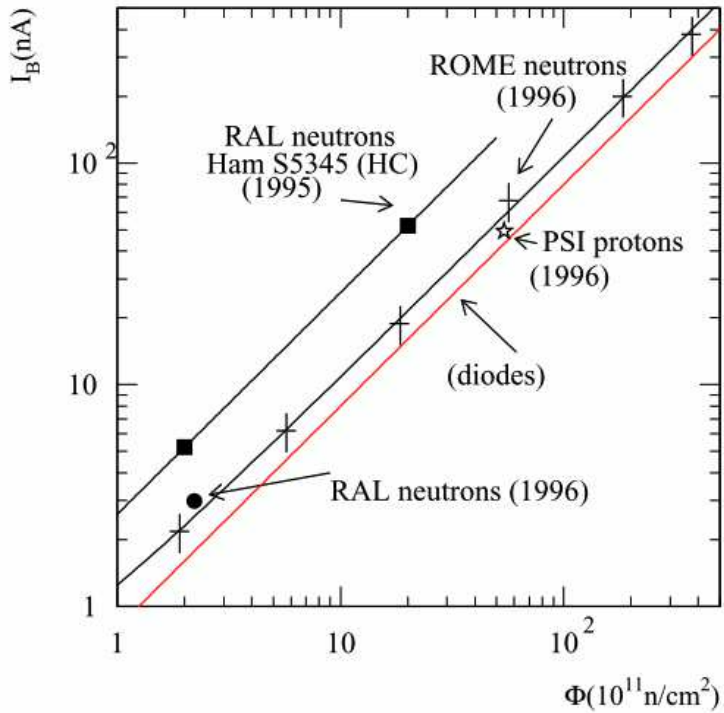}
\caption{Change in dark current for Hamamatsu APSs used in the electromagnetic calormeter of the CMS experiment, with respect to neutron fluence \cite{CMS}.}
\label{fig:apd}
\end{figure}

Scintillating crystal calorimeters are affected by TID and DDD in different ways. In inorganic scintillating crystals, ionizing radiation can create color centers which absorb scintillation light, reducing light yield. And in some cases also the scintillation centers could be damaged. Organic scintillators instead can also be damaged by DDD, especially by low energy neutrons which interact with hydrogen atoms modifying the molecular roto-vibrational levels and the related scintillation processes. TID of few Gray can already induce color centers in inorganic scintillators, depending on their structure \cite{zhu}. 
Photodetectors can be affected by TID in their optical coupling components, with photomultiplier tube photocathode windows being darkened by color center formation and silicon photodetectors having their optical resin damaged by DDD. Silicon bulk of solid state photodetectors can also be damaged by DDD and SEE as already described for silicon detectors \cite{CMS}. TID of the order of kGy can damage glass windows while neutron fluences of 10$^{11}$ can induce effects in silicon and optical resins.

\section{Radiation damage in space environment}

Space environment is rich in different types of radiation with a wide energy spectrum. Solar wind and exceptional solar events, like storms or flares on its surface, continuously stream electrons and protons towards the Earth. Sources outside the solar system also generate radiation with different wavelength, and high energy particles up to 10$^{21}$ eV.
Most of the charged particles are trapped by the Earth magnetic field in the Van Allen belts and only few of them, the most energetic, reach the atmosphere where they interact generating air showers. Devices in space are thus exposed to different levels and kind of radiation depending on their orbit and their expected lifetime in space.

Radiation in space can be categorized as: predictable and unpredictable effects. 
The average ionizing radiation present in the Van Allen belts can be accounted for as a predictable amount of TID and DDD. These doses should be taken into account while designing electronic devices to be used on earth orbits, or outer space missions. The unpredictable part of space radiation includes exceptional radiation sources from solar flares or storms, and can cause SEE which can disturbe or permanently damage the exposed devices. Damage can go from transient software upsets, to data corruption and permanent hardware upset. Electronic components can remain functional after a TID from tens of Gy up to tens of kGy, depending on the technological properties \cite{space}.

\begin{figure}[h!]
\includegraphics  [width = 0.8  \textwidth] {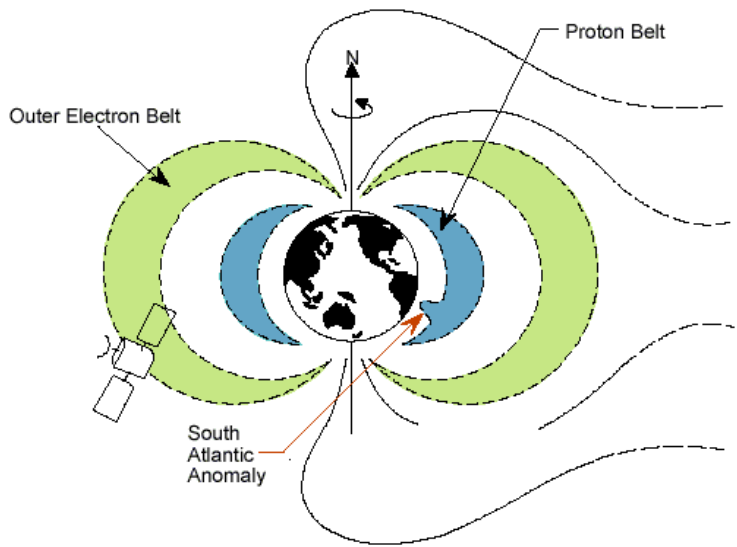}
\caption{Schematic view of charged particles trapped in the Van Allen belts by Earth magnetic field. The South Atlantic Anomaly is also visible.}
\label{fig:space}
\end{figure}

\section{Radiation damage in nuclear medicine}

\begin{figure}[h!]
\includegraphics  [width = 0.8  \textwidth] {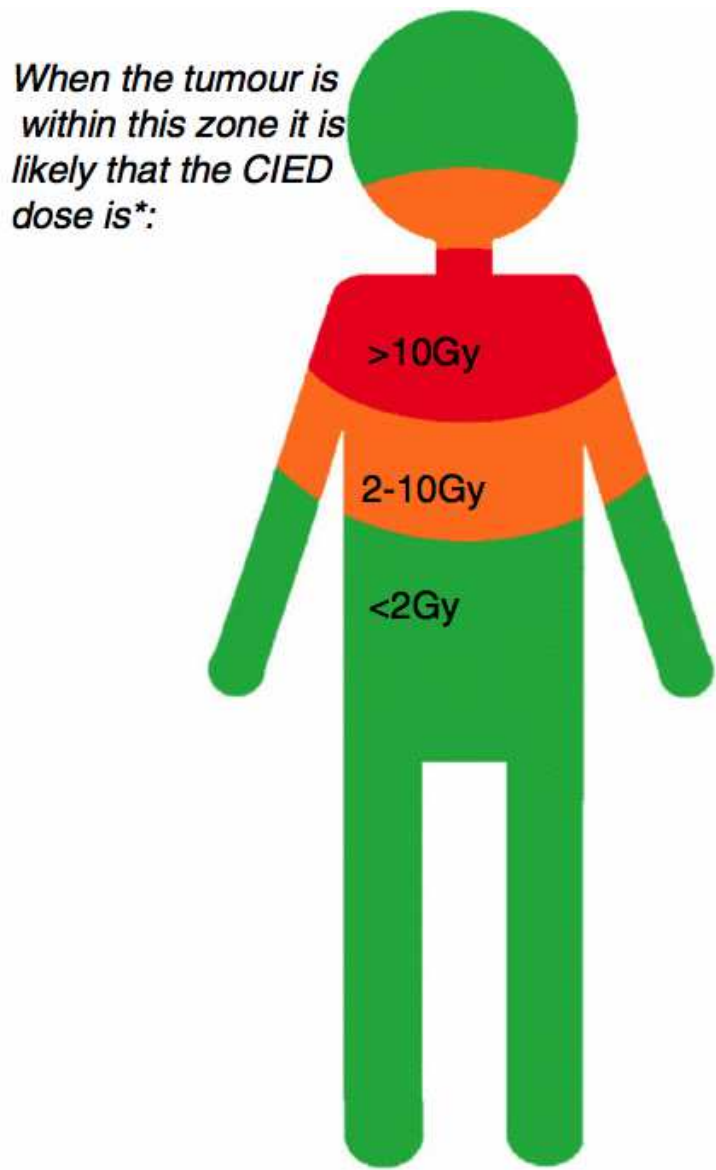}
\caption{Ionizing dose foreseen for CIEDs, when the patient is treated with radiotherapy in different areas of the body \cite{dutch}}
\label{fig:omino}
\end{figure}

The use of ionizing radiation in oncology has brought high radiation environments close to human body. Medical imaging with radiotracers exposes the patient to moderate levels of radiation depending on the radioisotope and the farmacokinetics of the binding molecule. A PET scan generally uses an activity of 18F-FDG equal to 10 mCi, which implies a dose delivery to the patient of 7 mSv \cite{ICRP}.
On the other hand the dose delivery to the PET scanning device would be lower due to the distance from the emission region located in the patient body, its screening effect and the lower exposure time. An average dose level to PET scanners has been estimated to be of the order of 0.05 mGy or less per single scan \cite{18f}. Therefore no damage to the active parts of the scanner is foreseen.
A different situation is present in radiotherapy treatments. Depending on the type of radiation, being gamma rays or protons or heavy ions, the area surrounding the target tissue could be interested by fringe radiation fields with an extension from few mm to few cm around the target. What can be damaged in these cases are implanted devices in the patient's body.
Implanted electronic devices include Cardiac Implanted Electronic Devices (CIED), Implanted Cardiac Defribillators (ICDs), Pace Makers, drug pumps, neurostimulators. These devices share the common trait to be often built with CMOS technology, which is sensitive to TID. After radiotherapy treatments these devices have reported to fail in different ways, for example  ICDs experienced from simple malfunctioning, like changes in intervention thresholds, up to complete failure  \cite{dutch}. During radiotherapy treatments such devices have a probability to receive a certain dose depending on the location of the device into the patient body (see fig.~\ref{fig:omino}). Three risk groups have been identified by the Dutch Society of Radiotherapy and Oncology for pacemakers and ICDs \cite{dutch}: a device exposed up to 2 Gy should only be checked for functionality after the treatment, exposure between 2 and 10 Gy needs the presence of a cardiologist and a crash cart during treatment, while over 10 Gy it should be discussed wether the radiotherapy or the cardiac disease are the main concern for the patient, and eventually relocate the device.

\section{Conclusions}
Detectors and electronic devices when used in radiation environment can undergo a number of different malfunctionings and ageing effects, depending on the technologies implemented in these devices. Thorough radation damage tests should be performed, simulating as much as possible the environmental radiation types and energies during real operation. This will allow the best technology to be chosen for the foreseen environment. Test facilities providing ionizing radiation (gamma, proton, ions), neutron research reactors and generators are available to meet these requirements for prototype irradiation. Test results should be integrated with Monte Carlo simulations of the radiation fields and beams together with experimental setups in order to better understand the effects on the exposed prototypes. A number of benefits would come from new radiation hard technologies, such as longer interplanetary missions with probes or manned missions, more reliable medical devices, and more detailed knowledge of the basic constituents of matter.

%
%

\end{document}